\documentstyle[prb,aps,twocolumn]{revtex}
\begin{document}
\draft
\title{Time-Varying Fine-Structure Constant Requires Cosmological 
Constant}
\author{Rainer W. K\"uhne}
\address{Vorm Holz 4, 42119 Wuppertal, Germany}

\maketitle

\begin{abstract}
Webb et al. presented preliminary evidence for a time-varying 
fine-structure constant. We show Teller's formula for this variation 
to be ruled out within the Einstein-de Sitter universe, however, 
it is compatible with cosmologies which require a large cosmological 
constant.
\end{abstract}

The possibility of time-varying physical constants was suggested by 
Dirac \cite{Dirac37,Dirac38} and Milne \cite{Milne}. This suggestion 
has been widely discussed 
\cite{Jordan,Teller,Dicke,Dyson,Carter,Carr,Ivanchik} for a number of 
motivations:

(i) The assumption of physical quantities being constant in space and 
time is ad hoc. It has to be either confirmed or rejected by 
observation \cite{Dirac37,Dirac38}.

(ii) The possibility of life as we know it depends on the values of a 
few basic physical constants and appears to be sensitive to their 
numerical values. This argument appeals to the anthropic principle 
\cite{Dicke,Carter,Carr}.

(iii) The numerical values of the basic dimensionless constants, e. g. 
the fine-structure constant, are not yet explained. If they depend 
on other constants and the age of the universe and are therefore 
time-varying, then the actual number of the independent free 
parameters of the standard theory of particle physics would be reduced.

Products and quotients of physical constants have been shown to be of 
fundamental importance for new physical phenomena:

(i) The unit of flux quantization in superconductors is $\phi =h/(2e)$, 
where $h$ denotes the Planck constant and $e$ the unit electric charge 
\cite{Deaver,Doll}. This finding was the first experimental proof of 
the famous BCS-theory \cite{Bardeen}. Furthermore, the flux quantization 
in units of $\phi$ is of central importance for the Josephson junctions 
\cite{Josephson}.

(ii) The characteristic magnetic flux in the 
Aharonov-Bohm effect \cite{Aharonov,Chambers} is 
quantized in units of $\phi =h/e$.

(iii) The unit conductance in the integer quantum Hall effect 
\cite{Klitzing} is $\hat\sigma = e^{2}/h$. In the fractional quantum 
Hall effect \cite{Tsui} the conductance is $\sigma =pe^{2}/h$, where 
$p$ is a rational number.

(iv) The Chandrasekhar limit mass \cite{Chandrasekhar} of white dwarf 
stars is
\begin{equation}
M_c = \frac{3\sqrt{\pi }}{2\mu^{2}M^{2}_p } \left( \frac{\hbar c}{G} 
\right)^{3/2},
\end{equation}
where $\hbar =h/(2\pi )$, $c$ is the speed of light, $G$ is Newton's 
gravitational constant, $M_p $ is the proton rest mass, and $\mu$ is 
the number ratio of nucleons and electrons.

In order to find further fundamental and important physical phenomena 
one is attempted to try to multiply or divide the Planck units 
\cite{Planck},
\begin{eqnarray}
l_p & = & (G\hbar /c^{3})^{1/2} \\
t_p & = & (G\hbar /c^{5})^{1/2} \\ 
m_p & = & (\hbar c/G)^{1/2},
\end{eqnarray}
with the Einstein constant \cite{Einstein},
\begin{equation}
\kappa =8\pi G/c^{4},
\end{equation}
the ``natural units'' $\hbar$ and $c$, and the present value $H_0$ of 
the Hubble parameter \cite{Hubble}. By just doing this, Teller 
\cite{Teller,Ivanchik} found the remarkable relation (in modern notation),
\begin{equation}
\kappa m_p H_0 c= \kappa\hbar H_0 /l_p =8\pi t_p H_0 
= \mbox{exp}(-1/ \alpha_0 ),
\end{equation}
where 
\begin{equation}
\alpha_0 = \frac{e^{2}}{4\pi\varepsilon_0 \hbar c}
\end{equation}
is the fine-structure constant. Here and in the following the subscript 
``0'' refers to the present value of the respective parameter.

From Teller's equation one can derive the change rate
\begin{equation}
\frac{\dot\alpha_z}{\alpha_z} = \alpha_z \frac{\dot H_z}{H_z} ,
\end{equation}
where the subscript ``$z$'' denotes the value of the respective 
parameter at cosmological redshift $z$ and the dot denotes the 
derivative with time. This rate can be determined experimentally. 
By examining quasar absorption lines in the redshift region 
$0.5 < z < 1.6$, Webb et al. \cite{Webb} found preliminary evidence for a 
time-varying fine-structure constant,
\begin{equation}
\frac{\Delta\alpha}{\alpha} = (-1.09\pm 0.36)\times 10^{-5},
\end{equation}
implying
\begin{equation}
\frac{\dot\alpha_z}{\alpha_z}= -1\times 10^{-5}H_z .
\end{equation}\label{Webb}
If the preliminary finding by Webb et al. will not be confirmed by 
future investigations, then Eq. 10 should be considered as 
an upper limit for the change rate of the fine-structure constant.

It is our aim to examine which cosmologies are compatible with both 
Teller's equation and (the upper limit of) the change rate of the 
fine-structure constant determined by Webb et al.

Compatibility is satisfied for cosmologies like the de Sitter 
universe \cite{deSitter} and the steady state cosmology \cite{Bondi,Hoyle}, 
where the Hubble parameter is constant in time. Unfortunately, these models 
have already been ruled out as viable cosmologies, because they contradict 
a number of other cosmological findings \cite{Peebles}.

According to general agreement the expansion of the universe is 
explained by the Friedmann-Lema\^itre equation 
\cite{Friedmann22,Friedmann24,Lemaitre},
\begin{equation}
H^{2}=\frac{8\pi G\varrho }{3} - \frac{kc^{2}}{R^{2}} 
+ \frac{\lambda c^{2}}{3}.
\end{equation}
The curvature radius $R$ and the mean mass density $\varrho$ scale 
with the redshift as,
\begin{eqnarray}
 R & = & R_0 /(1+z) \\
 \varrho & = & \varrho_0 (1+z)^{3}.
\end{eqnarray}
By defining the present values of the mass parameter
\begin{equation}
\Omega_0 = \frac{8\pi G \varrho_0}{3H^{2}_0}
\end{equation}
and the cosmological parameter
\begin{equation}
\lambda_0 = \frac{\lambda c^{2}}{3H^{2}_0},
\end{equation}
the Friedmann-Lema\^itre equation can be written as \cite{Hoell},
\begin{equation}
\left( \frac{H_z}{H_0} \right)^{2} = \Omega_0 (1+z)^{3} 
-( \Omega_0 + \lambda_0 -1)(1+z)^{2} + \lambda_0 .
\end{equation}

The Einstein-de Sitter universe \cite{EdS}, where $\Omega_0 =1$ and 
$\lambda_0 =0$, is a special case of the Friedmann-Lema\^itre 
universe. In this cosmology, Eq. 16 is reduced to
\begin{equation}
\left( \frac{H_z}{H_0} \right)^{2} = (1+z)^{3}.
\end{equation}
By using this equation Dyson \cite{Dyson,Ivanchik} derived from 
Teller's equation the formula
\begin{equation}
\alpha_z = \frac{\alpha_0}{1- \frac{3}{2}\alpha_0 \mbox{ln}(1+z)},
\end{equation}
which somewhat resembles expressions known from running coupling 
constants in quantum field theory. Dyson's equation (18)  
yields the rate
\begin{equation}
\frac{\dot\alpha_z}{\alpha_z}= - \frac{3}{2}\alpha_z H_z
\end{equation}
and is, unfortunately, ruled out by the experimental result  
of Webb et al. \cite{Webb}

By regarding Eq. 16 we find Teller's equation to be compatible 
with the observation by Webb et al. if the present value $\Omega_0$ of 
the mass parameter is extremely small and the cosmological parameter 
$\lambda_0$ slightly above unity.

The lower bound on $\Omega_0$ is given by the baryonic matter content 
of the universe. The visible baryonic matter of the universe was 
determined to be \cite{Persic}
\begin{equation}
\Omega_0 = 0.003h^{-1},
\end{equation}
where $h=H_0 /(100 \mbox{km/(s~Mpc)})$. (The parameter $h$ was introduced 
into cosmology for ease of notation; it should not be confused with the 
Planck constant.) This extremely small value of $\Omega_0$ is compatible 
with other determinations of $\Omega_0$ derived from examinations of the 
dynamics of clusters of galaxies if majority of their mass consists of 
a network of cosmic strings. This hypothetical network contributes 
to the local mass but do not influence the expansion of the universe 
\cite{Vilenkin}. Consequently, the mass contribution of the cosmic 
string network does not appear in the Friedmann-Lema\^itre equation. 
Furthermore, determinations of $\Omega_0$ from hot big bang 
nucleosynthesis are model-dependent (especially the degree of  
inhomogeneities in the early universe is not well understood), for a 
discussion see Refs. \onlinecite{Hata} and \onlinecite{Copi}.

If the observation by Webb et al. will be confirmed and Teller's 
equation is correct, then we can predict the following present values 
of cosmological parameters:
\begin{eqnarray}
H_0 & = & 69.7~\mbox{km/(s~Mpc)} \\
\Omega_0 & = & 0.004 \\
\lambda_0 & = & 1.004.
\end{eqnarray}
This value of $H_0$ is compatible with majority of its recent 
determinations, e. g. Pierce et al. \cite{Pierce} found 
$H_0 =(87\pm 7)~\mbox{km/(s~Mpc)}$, 
Freedman et al. \cite{Freedman} reported  
$H_0 =(80\pm 17)~\mbox{km/(s~Mpc)}$, and Tanvir et al. \cite{Tanvir} 
measured $H_0 =(69\pm 8)~\mbox{km/(s~Mpc)}$. Harris et al. \cite{Harris} 
found $H_0 =(77\pm 8) ~\mbox{km/(s~Mpc)}$ and Madore et al. \cite{Madore} 
reported $H_0 =(73\pm 21) ~\mbox{km/(s~Mpc)}$.

Interestingly, the cosmological parameters of the Bonn-Potsdam model  
\cite{Hoell,Liebscher1,Liebscher2} resemble those suggested above. 
Recently, van de Bruck \cite{Bruck1,Bruck2} has examined the effects of the 
hypothetical cosmic string network on the cosmological velocity field 
within the framework of this model and has shown them to be 
compatible with cosmological observations. 

The possible time-variation of the fine-structure constant implies the 
variation of at least one of the constants $c$, $\hbar$, $\varepsilon_{0}$, 
and $e$. A time-varying speed of light, for example, would imply a change in 
the radiation losses of accelerated charges, which would be of importance 
especially in the very early universe \cite{Ahluwalia}.

In practice, time-variations of quantities with nonzero dimension can be 
measured and defined only relative to other quantities which have the same 
dimension. We will suggest a convention where the most fundamental 
constants are defined as being constant in time and can serve as a 
reference for the possible time-variation of other quantities.

We suggest to consider the ``natural units'' $c$, $\hbar$, $k_B$, and 
$\varepsilon_0$ as fundamental and constant in time. This is because 
according to relativity, lengths are measured as time intervals 
multiplied with the speed of light. Even our unit length, the meter, 
is defined as the length which the speed of light covers in 1/299~792~458 
seconds. Furthermore, quantum phenomena are described either by the 
particle picture or the wave picture, respectively. The ``conversion 
factor'' of some of the quantities of this duality is $\hbar$. Finally, 
temperature is a macroscopic quantity and can be defined by the energy 
of microscopic objects divided by the Boltzmann constant $k_B$.

According to concepts of quantum gravity, the Planck units, especially 
the Planck time, are most fundamental. Hence, we suggest to measure all 
quantities in units of the Planck units (and multiples thereof), 
$t_p$, $l_p = ct_p$, $m_p = \hbar /(cl_p )$, $T_p = m_p c^{2}/k_B $, 
and $I_p = ( \varepsilon_0 \hbar c)^{1/2} /t_p$. Under these conventions, 
any time-variation of the fine-structure constant implies the 
variation of the unit electric charge $e$.

\end{document}